\documentclass[aps,prl,nofootinbib,twocolumn,preprintnumbers]{revtex4-1}
\usepackage{amsmath,amssymb,graphicx,booktabs,bm,psfrag}
\pdfoutput=1
\usepackage{color}
\newcommand{\spac}{{\hspace{0.3mm}}}

\newcommand{\circdot}{\circ}
\newcommand{\cusp}{c}
\newcommand{\im}{G}

\begin{document}

\title{Resummation of Super-Leading Logarithms}

\preprint{MITP-21-033}
\preprint{July 2, 2021}

\author{Thomas Becher$^a$}\email{becher@itp.unibe.ch}
\author{Matthias Neubert$^{b,c}$}\email{matthias.neubert@uni-mainz.de}
\author{Ding Yu Shao$^{d\,}$}\email{dingyu.shao@cern.ch}

\affiliation{${}^a$Institut f\"ur Theoretische Physik {\em \&} AEC, Universit\"at Bern, Sidlerstrasse 5, CH-3012 Bern, Switzerland\\
${}^b$PRISMA$^+$ Cluster of Excellence {\em \&} MITP, Johannes Gutenberg University, 55099 Mainz, Germany\\
${}^c$Department of Physics, LEPP, Cornell University, Ithaca, NY 14853, U.S.A. \\
${}^d$Department of Physics, Center for Field Theory and Particle Physics {\em \&} Key Laboratory of Nuclear Physics and Ion-beam Application (MOE), Fudan University, Shanghai, 200433, China}

\begin{abstract}
Jet cross sections at high-energy colliders  exhibit intricate patterns of logarithmically enhanced higher-order corrections. In particular, so-called non-global logarithms emerge from soft radiation emitted off energetic partons inside jets. While this is a single-logarithmic effect at lepton colliders, at hadron colliders phase factors in the amplitudes lead to double-logarithmic corrections starting at four-loop order. This effect was discovered a long time ago, but not much is known about the higher-order behavior of these terms and their process dependence. We derive, for the first time, the all-order structure of these ``super-leading logarithms'' for generic $2\to l$ scattering processes at hadron colliders and resum them in closed form.
\end{abstract}
\maketitle

If the radiation in a high-energy scattering process is restricted by experimental cuts, higher-order terms in the perturbative series are enhanced by large logarithms associated with soft and collinear emissions. The simple structure of these emissions often makes it possible to resum the logarithmic terms to all orders, either analytically or using parton-shower methods. For non-global observables, such as exclusive jet cross sections in which a veto on radiation is imposed only in certain angular regions, even the leading logarithms have a complicated structure due to the fact that they are generated by secondary emissions off the original hard partons \cite{Dasgupta:2001sh}. 

The prototypical non-global observable is the interjet energy flow, where a veto associated with a low scale $Q_0$ is imposed on radiation in a region away from the hard jets with energy of the order of the collision energy $Q$. Being sensitive only to large-angle soft radiation, one expects the leading logarithms to this observable to scale as $\alpha_s^n\spac L^n$, where $L=\ln(Q/Q_0)$. This is indeed the case for $e^+ e^-$ colliders, but Forshaw, Kyrieleis and Seymour \cite{Forshaw:2006fk} argued that at hadron colliders double logarithms arise at four-loop order, so that the leading logarithm at this order is $\alpha_s^4 L^5$. These so-called super-leading logarithms (SLLs) are a subtle effect generated by complex phases in the amplitudes, which spoil the real-virtual cancellation for collinear emissions off the initial states \cite{Catani:2011st,Forshaw:2012bi,Schwartz:2017nmr}. The effect is absent in the large-$N_c$ limit and not captured by any of the existing parton showers, which therefore do not account for the leading-logarithmic corrections to non-global observables at hadron colliders.

Even 15 years after this effect was discovered, remarkably little is known about it. While the first SLL is known for arbitrary $2\to2$ hard processes \cite{Forshaw:2008cq}, the second SLL ($\sim\alpha_s^5\spac L^7$) is known for some selected partonic channels only \cite{Keates:2009dn}. The all-order structure of SLLs, their contribution to other hard processes and their large-order behavior are completely unknown. One reason for this lack of understanding lies in the fact that one needs to perform calculations in the full color space, whose dimension is rapidly growing with the number of emitted partons. 

In \cite{Becher:2015hka,Becher:2016mmh} we have derived factorization theorems for non-global observables in Soft-Collinear Effective Theory (SCET) \cite{Bauer:2000yr,Bauer:2001yt,Beneke:2002ph} and found that non-global logarithms are governed by a renormalization-group (RG) equation. Here we apply this method to non-global logarithms at hadron colliders and derive the all-order structure of the SLLs $\alpha_s^{3} L^3\times  \alpha_s^{n} L^{2n}$ for arbitrary $2\to l$ processes. We further show that the effect already arises  for $l=0$, relevant e.g.\ to Higgs production with a central jet veto. 

As a concrete example, we consider the $pp \to 2~\text{jet}$ cross section  with a veto on hard radiation in a rapidity region $\Delta Y$ in between the two leading jets. This can  be imposed by requiring that any additional jet in the veto region has a transverse momentum smaller than $Q_0$. At leading logarithmic accuracy, there is no sensitivity to how the radiation is vetoed but only to the scale hierarchy between $Q_0$ and the transverse momentum of the hard jets, which is of order the partonic center-of-mass energy,  $Q=\sqrt{\hat{s}}=\sqrt{x_1 x_2 s}$. For this ``gap between jets'' observable, the following factorization formula holds \cite{Balsiger:2018ezi}: 
\begin{equation}\label{hadronfact}
\begin{aligned}
   &\sigma(Q_0) = \sum_{a_1,a_2=q,\bar{q},g} \int d x_1 d x_2 \\
   &\times \sum_{m=4}^\infty \big\langle \bm{\mathcal{H}}_m(\{\underline{n}\},Q,\mu)
    \otimes \bm{\mathcal{W}}_m(\{\underline{n}\},Q_0,x_1,x_2,\mu) \big\rangle \,.
\end{aligned}
\end{equation}
The hard functions $\bm{\mathcal{H}}_m$ describe all possible  $m$-parton processes $a_1+a_2\to a_3+\dots+a_m$ and are obtained after imposing appropriate kinematic constraints, such as cuts on the transverse momenta and rapidities of the leading jets. One then integrates over the phase space but for fixed directions $\{\underline{n}\}=\{n_1,\dots,n_m\}$ of the $m$ partons, i.e.\
\begin{equation}\label{eq:Hm}
\begin{aligned}
   \bm{\mathcal{H}}_m 
   &= \frac{1}{2\hat{s}} \prod_{i=3}^m \int\!\frac{dE_i\,E_i^{d-3}}{(2\pi)^{d-2}}\,
    |\mathcal{M}_m(\{\underline{p}\})\rangle \langle\mathcal{M}_m(\{\underline{p}\})| 
    \\[1mm]
   &\times (2\pi)^d\,\delta(\sqrt{\hat{s}}-E_{\rm tot})\,
    \delta^{(d-1)}(\vec{p}_{\rm tot})\,\Theta_{\rm hard}\!
    \left(\left\{\underline{p}\right\}\right) ,
\end{aligned}
\end{equation}
where $E_{\rm tot}$ and $\vec{p}_{\rm tot}$ are the total energy and momentum of the final-state particles in the partonic center-of-mass frame. Note that the amplitude is squared in the sense of a density matrix. We use the color-space formalism \cite{Catani:1996vz}, and the color indices of the amplitude $|\mathcal{M}_m(\{\underline{p}\}) \rangle$ and its conjugate are not contracted. The color sum, indicated by $\langle\dots\rangle$ in \eqref{hadronfact}, is performed after the hard function is combined with the function $\bm{\mathcal{W}}_m$, which encodes the soft and collinear low-energy dynamics. Both quantities depend on the directions $\{\underline{n}\}$ of the hard partons, and after combining them the integrals over these directions are performed, as indicated by the symbol $\otimes$.

The function $\Theta_{\rm hard}$ enforces the constraints on the hard jets and ensures that no hard radiation enters the veto region. For the validity of formula \eqref{hadronfact} it is important that these constraints are compatible with factorization. The low-energy matrix elements $\bm{\mathcal{W}}_m$ consist of squared  matrix elements of $m$ soft Wilson lines for the incoming and outgoing partons together with two collinear fields for the incoming particles. They need to be evaluated in SCET with Glauber gluons \cite{Rothstein:2016bsq}, which can mediate nontrivial interactions between soft and collinear partons. The functions $\bm{\mathcal{W}}_m$ contain rapidity logarithms, which induce a logarithmic dependence on the scale ratio $\sqrt{\hat{s}}/Q_0$ \cite{Becher:2010tm,Chiu:2012ir}. It would be interesting to analyze the structure of these matrix elements in more detail in future work. Here we just note that the additional dependence on the hard scale is single logarithmic, while we focus on the leading double-logarithmic corrections in this Letter.

To obtain the leading double logarithms, we solve the RG equation for the hard function iteratively and evolve it from the hard scale $\mu_h\sim\sqrt{\hat{s}}$ to the low scale $\mu_s\sim Q_0$. As the starting point of the evolution we use the lowest-order (Born level) hard function, which for a two-jet cross section involves four partons. We thus evaluate 
\begin{align}\label{eq:US}
   &\bm{\mathcal{H}}_4(\mu_h)\,\bm{U}(\mu_h,\mu_s) 
    = \bm{\mathcal{H}}_4(\mu_h)\,{\bf P}\exp\left[ 
    \int_{\mu_s}^{\mu_h}\!\frac{d\mu}{\mu}\,\bm{\Gamma}_H(\mu) \right] \nonumber\\
   &= \bm{\mathcal{H}}_4(\mu_h) + \int_{\mu_s}^{\mu_h}\!\frac{d\mu}{\mu}\,
    \bm{\mathcal{H}}_4(\mu_h)\,\bm{\Gamma}_H(\mu) \\
   &\quad + \int_{\mu_s}^{\mu_h}\!\frac{d\mu}{\mu}\,
    \int_{\mu}^{\mu_h}\!\frac{d\mu'}{\mu'}\,\bm{\mathcal{H}}_4(\mu_h)\,
    \bm{\Gamma}_H(\mu')\,\bm{\Gamma}_H(\mu)+ \dots \,. \nonumber 
\end{align}
Below, we will identify the SLLs that arise in the products of anomalous dimensions and solve a recursion relation for them. As a final step, we compute the cross section in \eqref{hadronfact} using the lowest-order expression for $\bm{\mathcal{W}}_m$ at the low scale $\mu_s\sim Q_0$. At this order the soft Wilson lines are trivial and the collinear matrix elements reduce to the usual parton distribution functions, i.e.\
\begin{equation}
   \bm{\mathcal{W}}_m(\{\underline{n}\},Q_0,x_1,x_2,\mu_s) 
   = f_{a_1}(x_1)\,f_{a_2}(x_2)\,\bm{1} \,.
\end{equation}

The one-loop anomalous dimension matrix in \eqref{eq:US} can be split into two parts:  $\bm{\Gamma}_H = \Gamma_{C}\,\bm{1} + \bm{\Gamma}_{S}$. The first part concerns the purely collinear singularities and is present also for inclusive cross sections. It is given by the usual DGLAP kernels and involves a convolution over the momentum fractions of the incoming partons. The second part, $\bm{\Gamma}_{S}$, contains soft as well as soft$\,+\,$collinear terms. This part is absent for inclusive cross sections, but present in our case because of the restrictions on hard radiation in the veto region. The soft$\,+\,$collinear piece generates the SLLs. The soft part of the anomalous dimension takes the form \cite{Becher:2015hka,Becher:2016mmh}
\begin{equation}\label{eq:gammaOne}
   \bm{\Gamma}_{S} = \frac{\alpha_s}{4\pi} \left(
    \begin{array}{ccccc}
     \bm{V}_{4} & \bm{R}_{4} & 0 & 0 & \cdots \\
     0 & \bm{V}_{5} & \bm{R}_{5} & 0 & \cdots \\
     0 & 0 & \bm{V}_{6} & \bm{R}_{6} & \cdots \\
     0 & 0 & 0 & \bm{V}_{7} & \cdots \\
     \vdots & \vdots & \vdots & \vdots & \ddots \\
    \end{array} \right)  + \dots\,.
\end{equation}
The virtual contributions $\bm{V}_{m}$ leave the number of partons unchanged, while the real-emission operators $\bm{R}_{m}$ add one extra parton to a given hard function. 

\begin{figure}[t!]
\centering
\begin{tabular}{cc}
\includegraphics[width=0.47\textwidth]{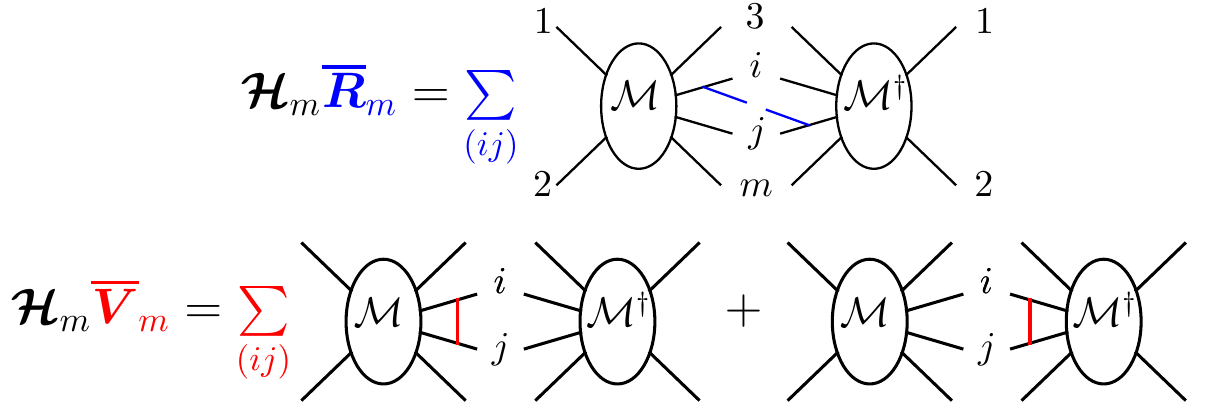} 
\end{tabular}
\caption{Action of the real-emission operator $\bm{\overline{R}}_m$ and the virtual piece $\overline{\bm{V}}\!_m$ on a hard function $\bm{\mathcal{H}}_m$.  Due to the emitted gluon (blue), the product $\bm{\mathcal{H}}_m\,\bm{\overline{R}}_m$ defines a hard function with $(m+1)$ external legs.}
\label{fig:VR}
\end{figure}

Due to the correspondence between UV and IR singularities \cite{Becher:2009qa}, the anomalous dimension $\bm{\Gamma}_S$ can be extracted by considering soft limits of hard-scattering amplitudes \cite{Becher:2016mmh,Balsiger:2018ezi}. For the present discussion, it is useful to write it in the form \cite{inprep}
\begin{equation}\label{eq:gammaSimp}
\begin{aligned}
   \bm{V}_m &= \overline{\bm{V}}\!_m + \bm{V}^{\im} + \sum_{i=1,2} \bm{V}^\cusp_i\,
    \ln\frac{\mu^2}{\hat{s}} \,, \\
   \bm{R}_m &= \overline{\bm{R}}_m + \sum_{i=1,2} \bm{R}^\cusp_i\,
    \ln\frac{\mu^2}{\hat{s}} \,,
\end{aligned}
\end{equation}
with 
\begin{align}
   \overline{\bm{V}}\!_m 
   &= 2\spac\sum_{(ij)}\,\big( \bm{T}_{i,L}\cdot\bm{T}_{j,L} 
    + \bm{T}_{i,R}\cdot\bm{T}_{j,R} \big)
    \int\frac{d\Omega(n_k)}{4\pi}\,\spac\overline{W}_{ij}^k \,, \nonumber\\
   \bm{V}^\cusp_i 
   &= 4\spac C_i\,\bm{1} \,, \nonumber\\[2mm]
   \bm{V}^{\im}
   & = - 8\spac i\pi\,\big( \bm{T}_{1,L}\cdot\bm{T}_{2,L}
    - \bm{T}_{1,R}\cdot\bm{T}_{2,R} \big) \,, \\[1mm]
   \overline{\bm{R}}_m 
   &= - 4\spac\sum_{(ij)}\,\bm{T}_{i,L}\circdot\bm{T}_{j,R}\,\spac
    \overline{W}^{m+1}_{ij}\,\Theta_{\rm hard}(n_{m+1}) \,, \nonumber\\
   \bm{R}^\cusp_i 
   &= - 4\spac\bm{T}_{i,L}\circdot\bm{T}_{i,R}\,\delta(n_k-n_i) \,. \nonumber
\end{align}
Before discussing the different parts in detail, let us explain how they act on a generic hard function $\bm{\mathcal{H}}_m$. The color generators $\bm{T}_{i,L}$ act on the amplitude and hence multiply $\bm{\mathcal{H}}_m$ from the left, while the generators $\bm{T}_{j,R}$ act on the conjugate amplitude and stand on the right of $\bm{\mathcal{H}}_m$. The color matrices in the virtual part act on the color indices of the $m$ partons, $\bm{T}_{i}\cdot\bm{T}_{j}=\sum_a\bm{T}_i^a\spac\bm{T}_j^a$, and $\bm{T}_i\cdot\bm{T}_i=C_i\,\bm{1}$ is the quadratic Casimir operator of parton $i$. This is the usual color-space notation. The color matrices in the real-emission terms $\bm{R}_m$ are different. They take an amplitude with $m$ partons and associated color indices and map it to an amplitude with $(m+1)$ partons, see Figure~\ref{fig:VR}. Explicitly, we have
\begin{equation}
   \bm{\mathcal{H}}_m\,\bm{T}_{i,L}\circdot\bm{T}_{j,R} 
   = \bm{T}_{i}^a\,\bm{\mathcal{H}}_m\,\bm{T}_{j}^{\tilde a} \,,
\end{equation}
where the color indices $a$ and $\tilde a$ refer to the emitted gluon. We use the symbol $\circdot$ to indicate the presence of the additional color space of the emitted parton. Subsequent applications of the anomalous-dimension matrix can act on these indices. In the simplest case of a single cut propagator as in Figure~\ref{fig:VR}, the indices are contracted with $\delta^{\tilde{a}a}$. On the other hand, if an additional gluon with group index $b$ is attached to the emitted parton, the indices get contracted with $(-if^{b\tilde{a}a})$.

The operators $\overline{\bm{V}}\!_m$ and $\overline{\bm{R}}_m$ encode soft singularities arising when a virtual or real soft parton is exchanged between two different legs of the hard function. The squared amplitude for the exchange is a product of the eikonal factors for each leg, and the notation $(ij)$ on the sums in \eqref{eq:gammaSimp} indicates a pair of unordered indices $i,j=1,\dots,m$. We use a bar to indicate that the collinear limits of the emissions are subtracted, i.e.\
\begin{equation}\label{eq:WijkPlus}
   \overline{W}_{ij}^k = \frac{n_i\cdot n_j}{n_i\cdot n_k\,n_j \cdot n_k} 
    - \frac{\delta(n_k-n_i)}{n_i\cdot n_k} - \frac{\delta(n_k-n_j)}{n_j\cdot n_k} \,.
\end{equation}
The angular $\delta$-distributions only act on the test function.

The collinear singularities in the soft anomalous dimension are encoded in  $\bm{R}^\cusp_i$ and $\bm{V}^\cusp_i$, both of which are proportional to the cusp anomalous dimension (as indicated by the superscript). These operators multiply a logarithm of the hard scale, which when inserted into \eqref{eq:US} gives rise to Sudakov double logarithms. We show below that all final-state collinear singularities cancel between real and virtual contributions, and for this reason only the initial-state pieces (with $i=1,2$) must be kept in \eqref{eq:gammaSimp}. The cancellation for the initial-state terms is spoiled by the complex Glauber phases in $\bm{V}^{\im}$, also referred to as Coulomb phases \cite{Forshaw:2006fk}. These arise whenever soft partons are exchanged between two final-state legs or two initial-state legs. Using color conservation, $\sum_{i=1}^m \bm{\mathcal{H}}_m\,\bm{T}^a_{i}=0$, the phase terms can be rewritten in the form of $\bm{V}^{\im}$, which makes it obvious that they are only relevant for processes involving (at least) two colored partons in the initial state.

Three properties of the different components of the anomalous dimension greatly simplify our calculations. Color coherence, the fact that the sum of the soft emissions off two collinear partons has the same effect as a single soft emission off the parent parton, implies that
\begin{equation}\label{eq:coherence}
   \bm{\mathcal{H}}_m\,\bm{\Gamma}^\cusp\,\bm{\overline{\Gamma}} 
   = \bm{\mathcal{H}}_m\,\bm{\overline{\Gamma}}\,\bm{\Gamma}^\cusp \,,
\end{equation}
where we have defined $\bm{\Gamma}^\cusp=\sum_{i=1}^2 (\bm{R}_i^\cusp+\bm{V}_i^\cusp)$ and $\bm{\mathcal{H}}_m\,\bm{\overline{\Gamma}}\equiv\bm{\mathcal{H}}_m \,(\bm{\overline{R}}_m+\bm{\overline{V}}\!_m)$. Next, the cyclicity of the trace ensures that
\begin{equation}\label{eq:collsafety} 
\begin{aligned}
   \left\langle \bm{\mathcal{H}}_m\,\bm{\Gamma}^\cusp\otimes\bm{1} \right\rangle 
   &= 0 \,, \\
   \left\langle \bm{\mathcal{H}}_m\,\bm{V}^{\im}\otimes\bm{1} \right\rangle 
   &= 0 \,.
\end{aligned}
\end{equation}
The first of these relations is a consequence of collinear safety: the singularity associated with a collinear real emission cancels against the one in the associated virtual correction. It is trivial to verify this, because
 \begin{equation}
   \left\langle \bm{\mathcal{H}}_m \left(\bm{R}_i^\cusp + \bm{V}_i^\cusp \right)
    \otimes \bm{1} \right\rangle 
   \propto \left\langle \bm{T}_i^a\,\bm{\mathcal{H}}_m\,\bm{T}_i^a
    - C_i\,\bm{\mathcal{H}}_m \right\rangle = 0 \,.
\end{equation}
The three properties hold for an arbitrary hard function $\bm{\mathcal{H}}_m$, which can be obtained from the tree-level hard function after applying the one-loop anomalous dimension several times. 

We extract the leading contributions to \eqref{eq:US} by considering products of $\bm{\Gamma}^\cusp$, $\bm{\overline{\Gamma}}$ and $\bm{V}^{\im}$, only the first of which gives rise to double logarithms. In the absence of $\bm{V}^{\im}$, we could use relation \eqref{eq:coherence} to move all occurrences of $\bm{\Gamma}^\cusp$ to the last step, where they give a vanishing contribution due to \eqref{eq:collsafety}. (Even in the presence of $\bm{V}^{\im}$ this can still be done for all final-state partons, and for this reason we did not include terms with $i\ne 1,2$ in the definition of $\bm{\Gamma}^\cusp$.) To get the SLLs, we thus need two insertions of $\bm{V}^{\im}$. A single insertion gives zero, since the cross section is real. Due to the two properties in \eqref{eq:collsafety} we also need one power of $\bm{\overline{\Gamma}}$ in the last step of the evolution. Therefore, the SLLs at $(3+n)^{\rm th}$ order in perturbation theory are associated with color traces of the form 
\begin{equation}\label{colortraces}
   C_{rn} = \big\langle \bm{\mathcal{H}}_4 \left(\bm{\Gamma}^\cusp\right)^r
    \bm{V}^{\im} \left(\bm{\Gamma}^\cusp\right)^{n-r} \bm{V}^{\im}\,\overline{\bm{\Gamma}} 
    \otimes\bm{1} \big\rangle\, ,
\end{equation}
where $0\le r\le n$. This explains why the SLLs first appear at four-loop order. However, the three-loop term $(n=0)$ originates from the same color structures and is numerically significant, even though it only involves the imaginary part $\pi =|\ln(-1)|$ of the large logarithm.

To get the corresponding contribution to the partonic cross section, we must combine the color traces $C_{rn}$ with the associated ordered integrals in \eqref{eq:US}. Each factor of $\bm{\Gamma}^\cusp$ is multiplied by a logarithm of $\mu$, see \eqref{eq:gammaSimp}, which produces a double logarithm upon integration. Neglecting the running of the coupling $\alpha_s$, setting $\mu_h^2 = \hat{s}$ and evaluating the integrals, we find with $L=\ln(\sqrt{\hat{s}}/\mu_s)$
\begin{equation}\label{eq:dsigma}
    \hat{\sigma}^{\rm SLL}_n = \left(\frac{\alpha_s}{4\pi}\right)^{n+3} L^{2n+3}\,
    \frac{(-4)^n\,n!}{(2n+3)!}\,\sum_{r=0}^n\,\frac{(2r)!}{4^r\spac(r!)^2}\,C_{rn} \,,
\end{equation}
which makes it explicit that starting from four-loop order two logarithms per loop arise.

The relations \eqref{eq:collsafety} imply that the color traces $C_{rn}$ can be simplified by working out the commutators $[\bm{V}^{\im},\overline{\bm{\Gamma}}\spac]$ and $[\bm{\Gamma}^\cusp,[\bm{V}^{\im},\overline{\bm{\Gamma}}\spac]]$. Under the trace, we find that both commutators evaluate to the same structure apart from a factor $(4N_c)$. We thus obtain
\begin{align}\label{step2}
   C_{rn} &= - 64\spac\pi \left( 4 N_c \right)^{n-r} f_{abc} \sum_{j>2}\spac 
    \big\langle\bm{{\cal H}}_4 \left( \bm{\Gamma}^\cusp \right)^r 
    \bm{V}^{\im}\spac\bm{T}_1^a\,\bm{T}_2^b\,\bm{T}_j^c \big\rangle \nonumber\\
   &\quad\times \int\frac{d\Omega(n_k)}{4\pi} \left( \overline{W}_{1j}^k
    - \overline{W}_{2j}^k \right) \Theta_{\rm veto}(n_k) \,.
\end{align}
The sum over $j$ contains the final-state partons of the Born process and the collinear gluons emitted from the $r$ remaining insertions of $\bm{\Gamma}^\cusp$, but not the initial-state partons 1 and 2. The contributions where $j$ refers to one of the collinear gluons emitted from the first $(n-r)$ insertions of $\bm{\Gamma}^\cusp$ in \eqref{colortraces} vanish. The gluon with label $k$ originates from the insertion of $\overline{\bm{\Gamma}}$ and must be attached to one initial-state and one final-state parton. The constraint $\Theta_{\rm veto}(n_k)=1-\Theta_{\rm hard}(n_k)$ restricts the emission to the veto region and arises from the incomplete cancellation of real and virtual terms in $\overline{\bm{\Gamma}}$. Since the direction $n_k$ in \eqref{step2} is in the veto region, it cannot be collinear to the directions $n_1$, $n_2$ or $n_j$. As a consequence, the collinear subtraction terms in \eqref{eq:WijkPlus} vanish, and one can replace $\overline{W}_{ij}^k \to W_{ij}^k$ in \eqref{step2}. 

All information about the phase-space restrictions on the direction of parton $k$ are contained in the angular integrals
\begin{align}\label{eq:Jints}
   J_j &= \int\frac{d\Omega(n_k)}{4\pi} \left( W_{1j}^k
    - W_{2j}^k \right) \Theta_{\rm veto}(n_k) \,.
\end{align}
The parton $j$ can either move along the directions $n_1$ and $n_2$, when it is attached to one of the collinear gluons emitted by the insertions of $\bm{R}^\cusp_i$, or it is one of the final-state partons. Since $W_{ii}^k$ vanishes we have $J_1=-J_2$. There are thus $(l+1)$ independent kinematic structures for a $2\to l$ jet process. For the gap between jets case, we find that $J_j =+\Delta Y$ if the rapidities of particles $j$ and 1 have opposite signs, and $J_j =-\Delta Y$ otherwise.

A more complicated structure arises when one commutes the remaining insertion of $\bm{V}^{\im}$ in \eqref{step2} all the way to the right. This leads to an expression involving anti-commutators of color generators, which in general cannot be simplified using the Lie algebra of $SU(N_c)$. 
Here we consider the important special case where particles 1 and 2 transform in the fundamental representation. We can then use the relation 
\begin{equation}\label{eq:dsymb}
   \{ \bm{T}_i^a,\bm{T}_i^b \}
   = \frac{1}{N_c}\,\delta_{ab}\,\bm{1} + \sigma_i\,d_{abc}\,\bm{T}_i^c \,; 
    \quad i=1,2 \,,
\end{equation}
where the color-space formalism implies that $\sigma_i=1$ for an initial-state anti-quark and $\sigma_i=-1$ for an initial-state quark. In this case a closed expression for the color traces $C_{rn}$ can be obtained, which involves only three non-trivial color structures:
\begin{align}\label{fundrep}
   C_{rn} &=2^{8-r}\pi^2 \left( 4 N_c \right)^n 
    \bigg\{ \sum_{j>2}\spac J_j\,\big\langle \bm{{\cal H}}_4 \big[ (\bm{T}_2-\bm{T}_1)\cdot\bm{T}_j \nonumber\\
  &\hspace{1.45cm}  + 2^{r-1} N_c \left( \sigma_1 - \sigma_2 \right) d_{abc}\,\bm{T}_1^a\,\bm{T}_2^b\,\bm{T}_j^c \big] 
    \big\rangle \\[1mm]
  &\quad + 2 \left( 1 - \delta_{r0} \right) J_{2}\, 
    \big\langle \bm{{\cal H}}_4\,\big[ C_F + \left( 2^r - 1 \right) \bm{T}_1\cdot\bm{T}_2 \big]
    \spac\big\rangle \bigg\} \,.\nonumber
\end{align}
The generalization of this result to the case of arbitrary representations involves a significantly larger number of color structures and will be discussed elsewhere \cite{inprep}. 

As a first application of the general result \eqref{fundrep} we consider quark-quark scattering. In this case the tree-level hard function has two possible color structures, octet or singlet, corresponding to gluon or photon exchange between the quarks. For the two cases, we get
\begin{align}
    C_{rn}^{(O)} 
    &= \hat{\sigma}_B\,2^{8-r}\pi^2 \left( 4N_c \right)^n 
     \bigg[ C_F J_{43} \nonumber\\
    &\hspace{1.55cm}+ \frac{J_2}{N_c} \left( N_c^2-2^{r+1}+1 \right) (1-\delta_{r0}) \bigg] \,, \\[1mm]
   C_{rn}^{(S)} 
   &= \hat{\sigma}_B\,2^{8-r}\pi^2 \left( 4N_c \right)^n C_F 
    \big[ - J_{43} + 2 J_2\,(1-\delta_{r0}) \big] \,, \nonumber
\end{align}
with $J_{43}=J_4-J_3$, and $\hat{\sigma}_B=\langle\bm{\mathcal{H}}_4\rangle$ is the Born-level partonic cross section. Assuming forward scattering as in \cite{Forshaw:2006fk}, the angular integrals evaluate to $J_2=J_{43}/2=\Delta Y$. Using these expressions in \eqref{eq:dsigma} and setting $n=1$, we recover the results of \cite{Forshaw:2006fk}. Repeating the calculation for $n=2$ we confirm the findings of \cite{Keates:2009dn}. As a further check of \eqref{fundrep}, we have written a computer code based on {\sc ColorMath} \cite{Sjodahl:2012nk} to directly evaluate the color structures $C_{rn}$ for fixed values of $r$ and $n$. Using this code, we have checked the general formula for $qq\to qq$,  $q\bar{q}\to q\bar{q}$ and $q\bar{q}\to gg$ scattering up to eight-loop order. 

The dependence of $C_{rn}$ in \eqref{fundrep} on $n$ and $r$ is powerlike, and it is possible to perform the sum over the infinite tower of SLLs in closed form:
\begin{equation}
    \Delta\hat{\sigma} = \sum_{n=0}^\infty \hat{\sigma}^{\rm SLL}_n 
    = \hat{\sigma}_B \left( \frac{\alpha_s}{4\pi} \right)^3 
     L^3\spac f(w) \,,
\end{equation}
where $w=\frac{N_c\alpha_s}{\pi}\,L^2$ encodes the double-logarithmic dependence. The function $f(w)$ can be expressed in terms of hypergeometric and related functions \cite{inprep}. For the singlet case, we get for forward scattering
\begin{equation}\label{eq:singletres}
   \Delta\hat\sigma^{(S)} 
   = - \hat{\sigma}_B\spac\frac{4C_F}{3\pi}\,\alpha_s^3\, L^3\spac
    \Delta Y\,{}_2F_2\big(1,1;2,\mbox{$\frac52$};-w\big) \,.
\end{equation}
While the explicit form is not particularly illuminating, it is interesting to study the asymptotic behavior for $w\to \infty$. Ordinary Sudakov double logarithms are resummed to the form $e^{-c\spac w}$ and are thus strongly suppressed in this limit, while the function $f(w) \sim (\ln w)/w$ falls off much slower.

\begin{figure}
\centering
\includegraphics[width=0.45\textwidth]{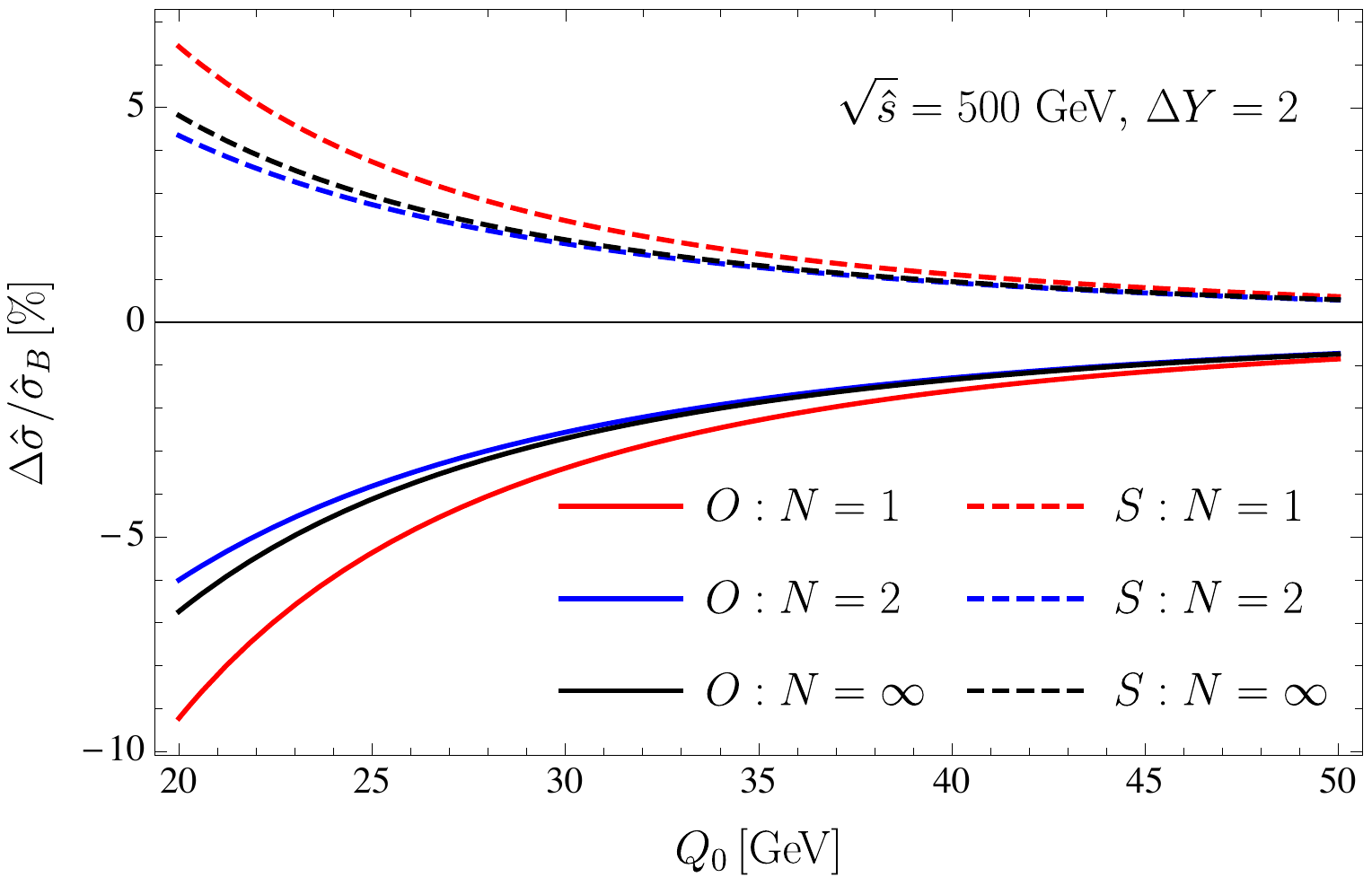} 
\caption{Super-leading logarithms in quark-quark scattering summed up to four-loop (red), five-loop (blue) and infinite order (black). The solid and dashed lines refer to the color octet and singlet channel, respectively.}
\label{fig:Kfac}
\end{figure}

In Figure \ref{fig:Kfac}, we evaluate the partonic $qq\to qq$ scattering cross sections for the octet and singlet channels. In order to only show the effect of SLLs, we plot the partial sums $\sum_{n=1}^N \hat{\sigma}_n^{\rm SLL}$ for different values of $N$. This omits the three-loop contribution from $\Delta\hat\sigma$, but note that also this term is due to complex phases not captured in conventional parton showers, see e.g.\ \cite{Nagy:2019rwb}. Due to the high power of $\alpha_s$, the SLLs are only significant if the logarithms are sizeable, and their effect is quite sensitive to the choice of scale in $\alpha_s(\mu)$. In the plot we set $\mu=Q_0$.

So far we have discussed the case of $2\to 2$ scattering, but an analogous relation with $\bm{{\cal H}}_4$ replaced by $\bm{{\cal H}}_{2+l}$ holds for a (anti-)quark-initiated $2\to l$ jet process with $l\ge 0$. In particular, we find that SLLs also arise for processes with less than two final-state jets, a fact that has not been appreciated in the literature. For $2\to 0$ processes such as $q\bar q\to V$, where $V=\gamma,Z^0,W^\pm$ is a colorless boson, the sum over $j$ in \eqref{fundrep} is absent, and color conservation implies that 
\begin{equation}
   C_{rn} = - \hat{\sigma}_B\,2^{9-r}\pi^2\spac C_F 
    \left( 4 N_c \right)^n\!\left( 2^r - 2 \right)\!
    \left( 1 - \delta_{r0} \right) J_{2} \,,
\end{equation}
which vanishes for $n=1$. The SLLs therefore start at 5-loop order, one order higher than in the general case. For $2\to 1$ scattering processes such as $q\bar q\to V+\text{jet}$, the only term in the sum has $j=3$, and one can use color conservation to obtain
\begin{equation}
   C_{rn} =\hat{\sigma}_B\,2^{10-r} \spac\pi^2 \left( 4 N_c \right)^{n-1} \left( N_c^2 + 2^r - 2 \right)
    \left( 1 - \delta_{r0} \right) J_{2} \,.
\end{equation}
These contributions start at four-loop order. In the literature \cite{Forshaw:2006fk,Keates:2009dn}, it has been stated that SLLs only arise when there are at least two colored partons in the final state, but as we have shown the emission into the gap originating from $\overline{\bm{\Gamma}}$ supplies the necessary additional parton for the $2\to 1$ case. For $2\to 0$ scattering the second final-state parton arises from a collinear emission in $\bm{\Gamma}^\cusp$, which explains why the effect is delayed by one order.

In this Letter we have solved the outstanding open problem of resumming SLLs for a large class of non-global observables at hadron colliders, thereby accounting for the leading logarithmic corrections to such processes for the first time. Our RG-based approach provides a transparent understanding of the underlying physics, and our analytical results should be useful in the ongoing effort to generalize parton showers to finite $N_c$, see e.g.\ \cite{Nagy:2019pjp,Hoche:2020pxj,Hamilton:2020rcu,DeAngelis:2020rvq}.
It will be interesting to perform a detailed analysis of SLLs for an observable such as the gap fraction, including the full set of partonic channels and accounting for running-coupling effects.  Our findings indicate that SLLs could have an appreciable effect on precision observables, in particular in Higgs production, where higher-order effects are generally large. Indeed, we find that the perturbative coefficients in gluon-induced $2\to 0$ processes are an order of magnitude larger than in the quark case studied here \cite{inprep}. 
\acknowledgements
\vspace{1mm}
\noindent
The research of T.B.\ is supported by the Swiss National Science Foundation (SNF) under grant 200020\_182038. The research of M.N.\ is supported by the Cluster of Excellence PRISMA$^+$ (EXC 2118/1) within the German Excellence Strategy (project ID 39083149). The research of D.Y.S.\ is supported by Shanghai Natural Science Foundation under Grant No. 21ZR1406100.

\end{document}